\documentclass[apjl, iop]{emulateapj}

\usepackage[normalem]{ulem}
\usepackage{color}
\newcommand  \mgii  {\ifmmode {\rm Mg}{\textsc{ii}} \else Mg\,{\sc ii}\fi}
\newcommand  \MGII  {\ifmmode {\rm Mg}\,{\sc ii}\,\lambda2798 \else Mg\,{\sc ii}\,$\lambda2798$\fi}
\newcommand  \siiv  {\ifmmode {\rm Si}\, {\sc iv}\ \else Si\,{\sc iv}\fi}
\newcommand  \SIIV  {\ifmmode {\rm Si}\,{\sc iv}\,\lambda1399 \else Si\,{\sc iv}\,$\lambda1399$\fi}
\newcommand  \cv  {\ifmmode {\rm C}\, {\sc v}\ \else C\,{\sc v}\fi}
\newcommand  \civ  {\ifmmode {\rm C}\, {\sc iv}\ \else C\,{\sc iv}\fi}
\newcommand  \ciii  {\ifmmode {\rm C}\, {\sc iii}\ \else C\,{\sc iii}\fi}
\newcommand  \CIV  {\ifmmode {\rm C}\,{\sc iv}\,\lambda1549 \else C\,{\sc iv}\,$\lambda1549$\fi}
\newcommand  \NV  {\ifmmode {\rm N}\,{\sc v}\,\lambda1240 \else N\,{\sc v}\,$\lambda1240$\fi}
\newcommand  \nv  {\ifmmode {\rm N}\,{\sc v}\ \else N\,{\sc v}\fi}
\newcommand  \nii  {\ifmmode \left[{\rm N}\,{\textsc ii}\right] \else [N\,{\sc ii}]\fi}
\newcommand  \oii  {\ifmmode \left[{\rm O}\,{\textsc ii}\right] \else [O\,{\sc ii}]\fi}
\newcommand  \oiii  {\ifmmode \left[{\rm O}\,{\textsc iii}\right] \else [O\,{\sc iii}]\fi}
\newcommand  \oiv    {\ifmmode \left[{\rm O}\,{\textsc iv}\right] \else O\,{\sc iv}\fi}
\newcommand  \ovi    {\ifmmode \left[{\rm O}\,{\textsc vi}\right] \else O\,{\sc vi}\fi}
\newcommand  \LyA  {\ifmmode {\rm Lyman}\,{\sc $\alpha$}\,\lambda1216 \else Lyman\,{\sc $\alpha$}\,$\lambda1216$\fi}

\newcommand  \lya {\ifmmode {\rm Ly$\alpha$}\ \else Ly{$\alpha$}\fi}
\newcommand  \feii     {Fe\,{\sc ii}}

\newcommand  \ALIII  {\ifmmode {\rm Al}\,{\sc iii}\,\lambda1857 \else Al\,{\sc iii}\,$\lambda1857$\fi}

\newcommand{\kms}{\ifmmode {\rm km\,s}^{-1} \else km\,s$^{-1}$ \fi}

\newcommand{\sii}{\mbox{{Si\,{\sc ii}}}}
\newcommand{\cii}{\mbox{{C\,{\sc ii}}}}

\newcommand  \hi     {H\,{\sc i}}

\shorttitle{Remarkable gradients in a quasar}
\shortauthors{Yi et al.}

\begin{document}


\title{    Radial  gradients  revealed by mutliscale outflows from down-the-barrel spectroscopy toward a quasar at redshift 3.4   } 
\author{Weimin Yi\altaffilmark{1},  Paola Rodr{\'\i}guez Hidalgo\altaffilmark{2},  Chen Chen\altaffilmark{3},  P. B. Hall\altaffilmark{4}, Zhicheng He\altaffilmark{5},  R. P. Easton\altaffilmark{2}, D.~P. Schneider\altaffilmark{6,7,8}, W.~N. Brandt\altaffilmark{6,7,8},   Xue-Bing Wu\altaffilmark{9,10}, Kai-Xing Lu\altaffilmark{1}, Chuanjun Wang\altaffilmark{1}, Yuanjie Feng\altaffilmark{1}  }

\altaffiltext{1}{Yunnan Observatories, Chinese Academy of Sciences, Kunming, 650216,  China} 
\altaffiltext{2}{Physical Sciences Division – School of STEM, University of Washington Bothell Bothell WA, 98011, USA}
\altaffiltext{3}{ Zhuhai College of Science and Technology,  Zhuhai 519000, China} 
\altaffiltext{4}{ Department of Physics and Astronomy, York University, Toronto, ON, M3J 1P3, Canada} 
\altaffiltext{5}{ CAS Key Laboratory for Research in Galaxies and Cosmology, Department of Astronomy, University of Science and Technology of China, Hefei, Anhui 230026, People’s Republic of China}
\altaffiltext{6}{Department of Astronomy \& Astrophysics, The Pennsylvania State University, 525 Davey Lab, University Park, PA 16802, USA}  
\altaffiltext{7}{Institute for Gravitation and the Cosmos, The Pennsylvania State University, University Park, PA 16802, USA}
\altaffiltext{8}{Department of Physics, 104 Davey Laboratory, The Pennsylvania State University, University Park, PA 16802, USA}  
\altaffiltext{9}{Kavli Institute for Astronomy and Astrophysics, Peking University, Beijing 100871, People’s Republic of China} 
\altaffiltext{10}{Department of Astronomy, Peking University, Yi He Yuan Lu 5, Hai Dian District, Beijing 100871, People’s Republic of China}

\begin{abstract}
Active galactic nucleus(AGN) feedback is a key ingredient in galaxy formation models and simulations. From an observational point of view,  however, the channels of AGN feedback coupling to  the interstellar medium (ISM) and  circumgalactic medium (CGM) and hence the impact on galaxy evolution,  are  largely uncertain and remain fiercely debated, due primarily to the huge gap from nuclear to CGM scales. Here we present multi-epoch, down-the-barrel spectroscopy toward a  luminous quasar at $z=3.409$ over two decades,  which reveals multiscale outflows expanding from nuclear to CGM scales along with characteristic radial gradients.  Most strikingly,  the trends of trough depth across three different-scale, freely expanding outflows are opposite between \nv\ and \civ, regardless of the spectral normalization and short-term variability, leading to a tenable gradient of N/C and signaling a critical  transition from  ejective  feedback on small scales to regulative feedback on large scales.  Our observations of this quasar offer valuable diagnostics to explore the realistic wind-ISM/CGM coupling,  one of the most challenging tasks  in state-of-the-art simulations of feedback.  
\end{abstract}


\section{Introduction}
\label{intro_sec}

Quasar winds or outflows, which are  traced primarily  by UV absorption  and/or emission  lines,  appear ubiquitous and signify ongoing feedback.  Generally,   there are three categories of UV winds based on their absorption trough width, 
namely  broad absorption line (BAL; \citealt{Weymann91}), mini-BAL (\citealt{Hamann01,Hamann13}), and narrow absorption line (NAL; \citealt{Foltz86,Vestergaard03}).  
In addition,  NALs with a  blueshift of $<3000$ \kms\ relative to the systemic redshift are generically termed associated absorption lines (AALs; \citealt{Foltz86}).    BALs are unambiguous quasar winds  driven by UV radiation from accretion disks (e.g., \citealp{Murray95,Proga00}), and  mini-BALs  appear to  resemble BALs in many aspects  that one can transform into the other  (e.g., \citealp{Paola13}) and exhibit  similar variability  (e.g., \citealp{Misawa14}).  
A growing number of extremely high-velocity outflows (EHVOs) in all the three aforementioned  outflow types  has been found  from quasar spectra (e.g., \citealp{Jannuzi96,Misawa07,Hamann13,Paola20,Aromal21,Vietri22}), signaling one of the most energetic  feedbacks.

BALs and mini-BALs appear  to be located in a broad range of distances from  the central engines  (from sub-pc to tens of kpc;  \citealp{Hamann01,Hamann13,Arav18,HeZ19,Byun22}).  Identifying the origin of an  NAL, however, is not trivial,  because  intrinsic NALs can be  mistakenly classified as  intervening gas if lacking the aid of high-resolution spectroscopy to check optical depth ratios or multi-epoch observations to test for variability (e.g., \citealp{Foltz86,Misawa07,Simon2010,ChenZ13b,Lewis2023,FuX2023}).  
 Although AALs are  likely to be  associated with quasars, they could form in a variety of environments, e.g., from circumnuclear ($\gtrsim$ a few pc) to intergalactic ($\gtrsim$ a few hundreds of kpc) scales (\citealp{Foltz86,Tripp1996,Wild2008,Prochaska2014,Perrotta2016,Lau2016,ChenC18,Culliton2019}).

Studies focusing on a single type of quasar winds (i.e., BAL, mini-BAL, NAL) are extensive (e.g., \citealp{Weymann91,Hamann01,Hamann13,ChenC18,Perna2025}), but there are not many published cases  based on the coexistence of BAL, mini-BAL,  NAL, and AAL  in the same spectrum  ( e.g., \citealp{Paola13,Misawa14}).  
From a statistical view,  intrinsic \civ\ NALs are often found to be associated with BALs and/or mini-BALs (e.g., \citealp{Stone19,Itoh20});  in particular, outflowing AALs appear to be strongly correlated with the presence of BALs (e.g., \citealp{ChenC20}), favoring the scenario where most AALs have formed in quasar-driven outflows, although the physical process remains debated  (e.g., \citealp{King11,Faucher12,Costa20,Faucher23}).

J075852.67+133530.8 (hereafter J0758) is a radio-quiet quasar that was archived  in the Sloan Digital Sky Survey (SDSS) Data Release 7 (DR7; \citealt{Schneider2010}).  It is one such unique case  possessing  distinct  UV winds with line-of-sight (LOS) velocities from 0 to  $\sim$40,000 \kms\ traced by blueshifted broad emission lines (BELs), a BAL, a mini-BAL, a NAL, and an AAL from nuclear to circumgalactic scales,  along with a BAL to non-BAL transformation over two decades, which offers a unique probe to bridge small- and large-scale outflows, one of the most challenging issues in  feedback.

%-------------------------------------------------------------
\begin{figure*}
        \centering % \flushleft
    \includegraphics[width=1.0\textwidth]{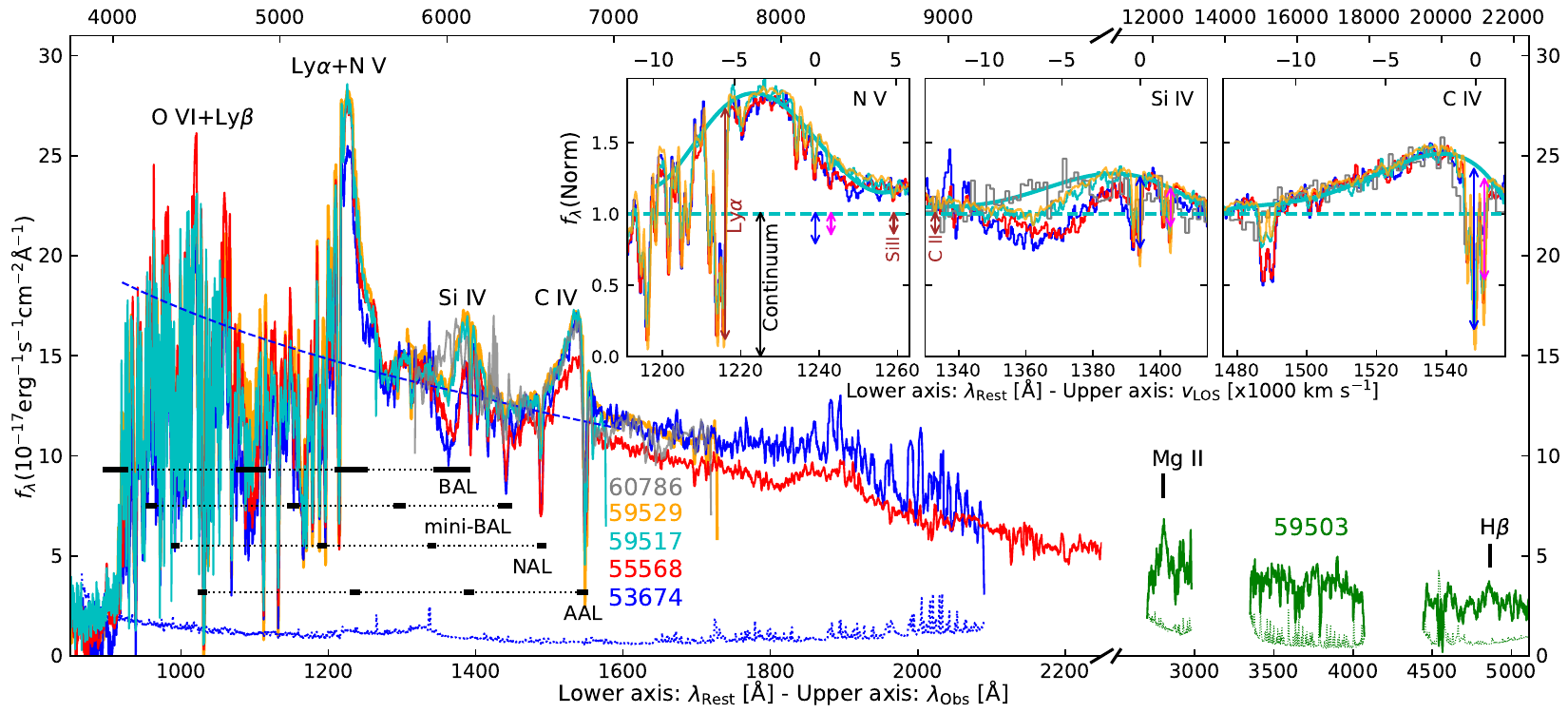}
     \caption{   The near-IR and multi-epoch optical spectra of J0758, in which the expected  positions and widths of the BAL, mini-BAL, NAL, and AAL are indicated  by  thick horizontal  bars  (each with \ovi, \lya+\nv, \siiv, and \civ\ from left to right).  The  blue dotted/dashed lines denote the spectral error/continuum fit to the first spectrum.  A systemic redshift of 3.409$\pm$0.001 was established from a combined analysis of  the \mgii\ and H$\beta$ broad emission lines.  Inset panels:  The continuum-normalized spectra,  in which the   thick lines denote  low-order polynomial fits to the \lya+\nv, \siiv, and \civ\  BELs  at MJD=59517 (after masking absorption features),  whose profiles are treated as the intrinsic BELs.  Singlet and doublet features in the AAL at MJD=53674 are indicated by the brown and blue/magenta arrows, respectively;  the arrows in \civ\ and Ly$\alpha$ have a length larger than that from the black arrow,  signaling a full coverage of the continuum.  
     } 
    \label{J075852_3spec_com}
\end{figure*}

\section{Observations}

We have obtained five  epochs of optical spectra and one near-IR spectrum for J0758.  Two of the optical spectra were obtained from public data release  of SDSS data (e.g., \citealp{York00,Eisenstein11}) and one from \citet{DESI2022}. Additional optical spectra were obtained by  the Low-Resolution Spectrograph-2 (LRS2; \citealt{Chonis16}) mounted on the Hobby-Eberly Telescope (HET; \citealt{Ramsey98,Hill21}),  with the spectroscopic data  processed by the LRS2 pipeline\footnote{https://github.com/grzeimann/Panacea}  which incorporates an improved procedure for flux calibration with a typical uncertainty of $\sim$15\%, and by the Yunnan Faint Object Spectrograph and Camera mounted on the Lijiang 2.4m Telescope (\citealt{WangC2019,WangJ2023}), with data reduction performed by IRAF packages via  standard data reduction steps including bias and sky subtractions, ﬂat-ﬁelding correction, and flux calibration.   Near-IR spectroscopic observations were performed with the TripleSpec spectrograph at the Palomar Hale 200 inch telescope (P200/TripleSpec; \citealt{Wilson04}). TripleSpec provides a wide wavelength coverage  (0.95--2.46 $\mu$m) at an average spectral resolution of $\sim2700$, allowing simultaneous observations in the J/H/K bands. A slit width of one arcsecond and the ABBA dither pattern along the slit were chosen to improve the sky subtraction during the observations.

The log of  observations  are summarized in Table.~\ref{table1} and the near-IR/optical spectra  are presented in Fig.~\ref{J075852_3spec_com} for an overall view. We highlight that  zero velocity is defined  by the shorter-wavelength component of a specific doublet at  systemic redshift.  The near-IR spectrum is clearly noisier  than those of the optical spectra,  so we  smoothed it by a 21-pixel Boxcar filter for visual inspection. All but the last-epoch optical spectra in Fig.~\ref{J075852_com_BAL_mini_NAL} are smoothed to an approximate resolution ($\sim$200 \kms) for consistency in comparison.   We measured a systemic redshift of 3.409$\pm$0.001 based on the \mgii\ and H$\beta$ broad emission lines observed in the near-IR spectrum.    A cosmology with  $H_0 = 70$~km~s$^{-1}$~Mpc$^{-1}$, $\Omega_M = 0.3,$ and $\Omega_{\Lambda} = 0.7$ is adopted throughout this work.

\begin{table}
\centering
 \caption{  Spectroscopic observations of J0758}
 \begin{tabular}{lcccc}
  \hline\noalign{\smallskip}
Instrument & $\lambda$/$\Delta\lambda$ &  Spectral  &  Exp  &  Observation  \\
Name &  &  Coverage  &   &  Date \\
           &  &  ($\mu$m)  &  (s)  &  (MJD) \\
  \hline\noalign{\smallskip}
SDSS & 1800 & 0.38--0.92 & 4800 & 53674 \\
SDSS & 1800 & 0.36--1.03 & 6300 & 55568 \\
P200/TripleSpec & 2700 & 1.0--2.5 & 1200 & 59503  \\
HET/LRS2 & 2500 & 0.36--0.7 & 1200 & 59517  \\
DESI  & 4000 & 0.36--0.96 & 3865 & 59529  \\
LJT/YFOSC & 300 & 0.65--0.9 & 2400 & 60786  \\
  \noalign{\smallskip}\hline
\end{tabular}
\label{table1}
\end{table}

\section{Data Analysis}

\subsection{The  Power-Law  Continuum Fit  and Reconstruction of the BELs}

We adopt a  power-law model to ﬁt the underlying UV continuum  based on the selection of common spectral regions  that are visually identified  to be  free of emission and absorption features  (1060--1070 \AA, 1280--1290~\AA, 1455–1470 \AA, 1570--1585 \AA).  The local continuum fit to the first-epoch spectrum  is  shown in Fig.~\ref{J075852_3spec_com}  for  visual  inspection.    The  rest-frame UV spectrum  is normalized by the corresponding  continuum fit  for each epoch.

As shown in the inset panels of Fig.~\ref{J075852_3spec_com}, the BEL profiles remain unchanged in \civ\ and Ly$\alpha$+\nv,  when variations  in the continuum are taken into  account.   
In contrast,  the blue wing of the  \siiv\ BEL, at first glance,  shows large variations from the spectra.  However, due to the presence of the \civ\ BAL (see \S\ref{trough_identification}) that overlaps with the \siiv\ BEL blue wing, one cannot draw a conclusion that the BAL  intrinsically varied  before  disentangling the potential contribution from BEL variability.  Since  \civ\ and \siiv\  have a comparable ionization potential, they are likely located at  the same or a nearby region,  which is supported by a similar kinematic profile between them.  Therefore,  it is reasonable to assume  that both the \civ\ and \siiv\ BELs remain unchanged, and  that the apparent change in the blue wing of the \siiv\ BEL is  due primarily to the \civ\ BAL variability.

Next, we attempt to reconstruct the  \siiv\ BEL,  despite the strong \civ\ absorption attached on it.  Fortunately, the last-epoch spectrum  provides a good benchmark to quantify the relative variability,  given its apparent lack of the \civ\ broad absorption;  thus, we chose a low-order polynomial function to fit the Ly$\alpha$+\nv, \siiv, and \civ\  BELs in the spectrum at MJD=59517 after masking  absorption regions.  We do not  tie the kinematics among the  three different-ion BELs during the fits due  to  potential  blending and mixing with other ions.  The  fitted  profiles are then used as  intrinsic BELs to match the observed BELs over the other spectroscopic epochs for each of them.   The \civ\ and \siiv\ BELs observed from the last-epoch spectrum  indeed  agree well with the reconstructed \siiv\ BEL,   although the spectral quality in the last epoch is too low to perform a reliable analysis of narrow absorption features.

To quantify the emission and absorption  strengths,  we measure the rest-frame equivalent widths (REW) from the continuum normalized spectra in each epoch, according to Equations (1) and (2) from \citet{Kaspi2002}. 
We consider two  independent errors when using Equation (2).  Specifically, the 1$\sigma$ spectral error distribution is retrieved  from the corresponding normalized spectrum,  and the uncertainty of the continuum placement  is calculated via a Monte Carlo approach following \citet{Yi19a}.  
Using  the same prescription of \citet{Filizak13}, we also calculated  the centroid velocity for the \civ\ BEL and each of the four absorption lines from the first-epoch spectrum at MJD=53674 (see Table~\ref{table2}).

\begin{figure*}
        \centering % \flushleft
        \includegraphics[width=1.0\textwidth]{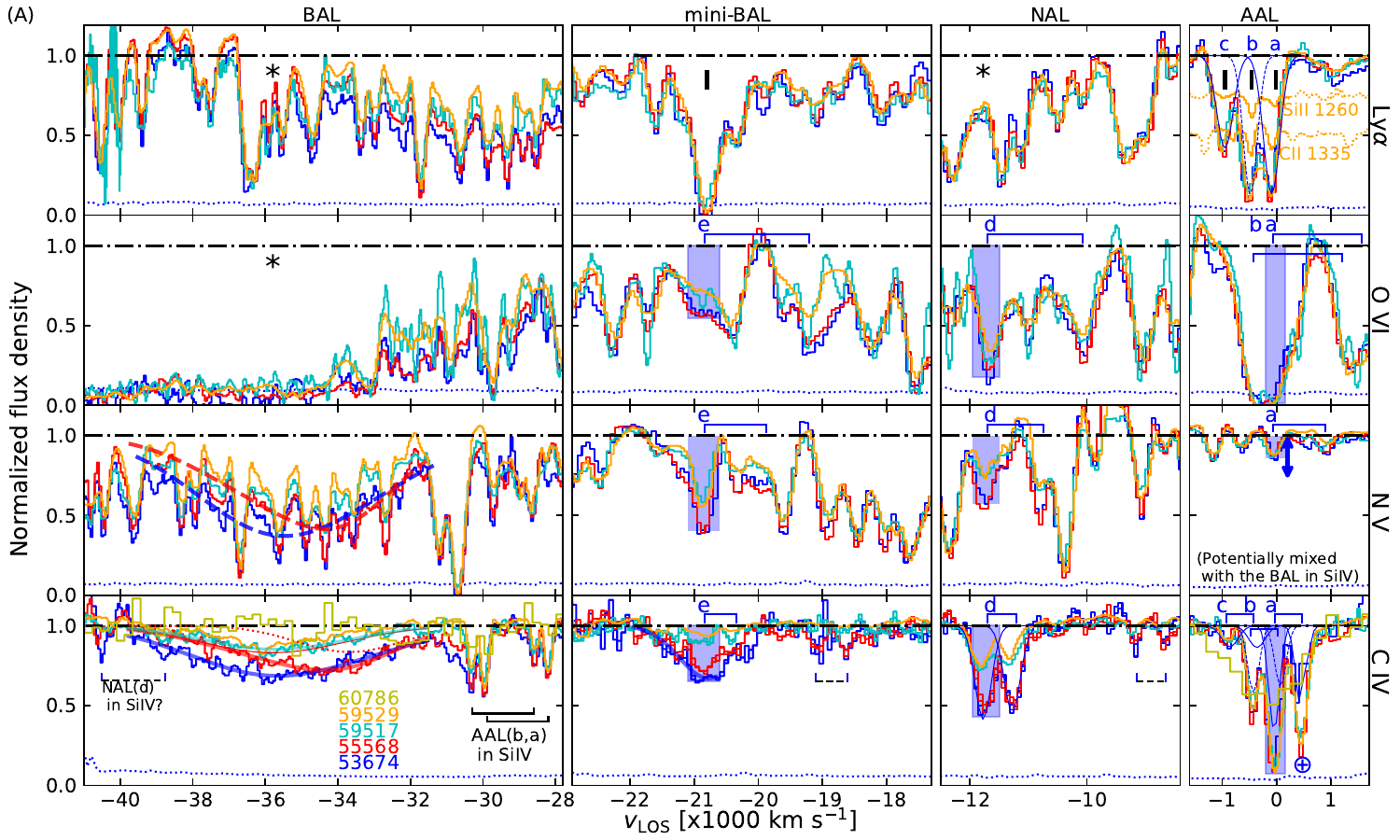}
         \includegraphics[width=.45\textwidth]{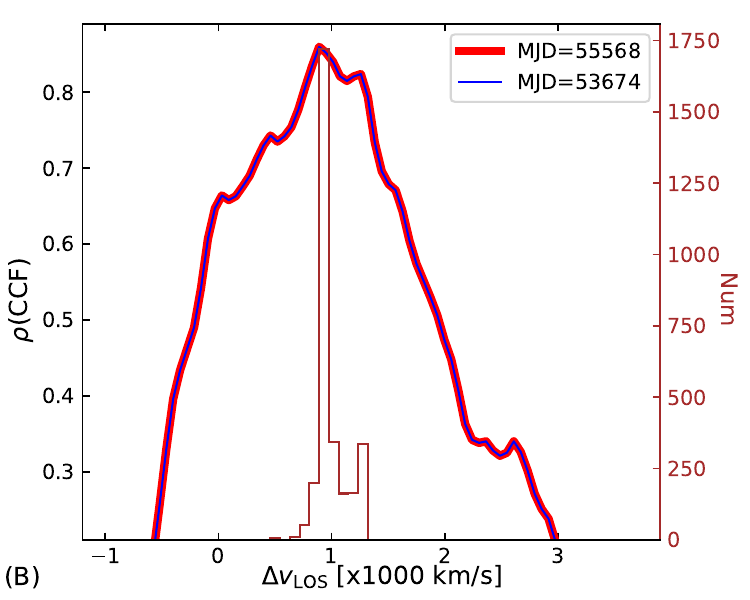}
         \includegraphics[width=.45\textwidth]{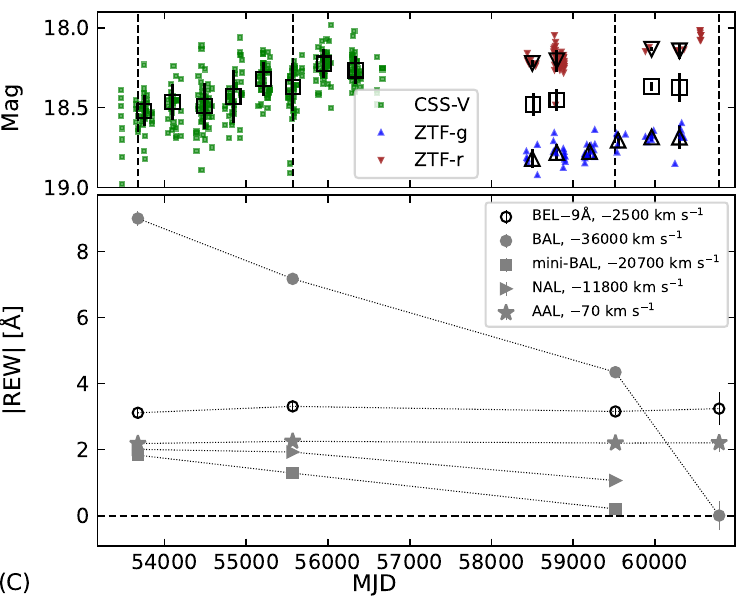}
    \caption{ 
     {\bf (A) } Demonstration of the different short-term variability behaviors and long-term trends across different-type absorption troughs  from the continuum+BEL normalized spectra.  The deepest portion of each \civ\ trough (last row) is used as the benchmark to identify the corresponding Ly$\alpha$, \ovi, and \nv\ troughs embedded in the Ly$\alpha$ forest from  the spectrum at MJD=53674. Thin/thick blue lines refer to the individual/total fits to these  troughs at MJD=53674;  the earth symbol is for telluric absorption; ''a,b,c,d,e'' for the identified  doublets;  dashed (inverted) hats  for tentative  ionic doublets;  ``*'' for  unidentifiable ionic troughs;  the blue dotted line for the spectral error at MJD=53674;  the arrow in the AAL-\nv\ panel for the upper limit on  ATD(\nv) derived from a normalization relative to the pure continuum.  The two fitted  \civ\ BAL profiles (MJD=53674 and 55568) broadly matches the corresponding \nv\ BAL troughs after an arbitrary scaling in flux. For the \civ-BAL trough at MJD=55568, at least two Gaussians (red solid/dotted  thin lines) are needed to fit it, in stark contrast with the \civ-BAL troughs modeled by one Gaussian over other epochs. The lowest S/N and resolution spectrum at MJD=60786 is displayed only in the \civ-BAL and \civ-AAL for visual inspection.  The highest-S/N spectrum at MJD=59529 reveals weak but significant absorption in \sii\ and \cii\ (shifted in Y-axis for clearity) from the AAL. Obviously, the ATD trends across the mini-BAL, NAL, and AAL outflows are opposite between \nv\ and \civ\ (or \ovi), regardless of the normalization level and short-term variability.    {\bf (B) } A monolithic shift of 950$_{-51}^{+126} \kms$ is measured via a CCF approach  across the entire \civ\ BAL trough from MJD=53674 to 55568,  suggestive of BAL deceleration  (see another possibility  in Section \ref{BAL_variability} for the BAL variability).    {\bf (C) }   Time variability of the \civ\ outflows in combination with the photometric light curve from CRTS (\citealp{Drake09}) and ZTF (\citealt{Masci19}), in which the ZTF-$r/g$ magnitudes have been converted to the $V$-band (open squares) using the equation from \citet{Jester2005} for easy comparison.   Note that the REW measurement is insensitive to spectral resolution. 
    }
    \label{J075852_com_BAL_mini_NAL}
\end{figure*}

\subsection{ Identification and quantification of  distinct \civ\  absorption features  }\label{trough_identification}

Based on the continuum+BEL normalized spectra  (see Fig.~\ref{J075852_com_BAL_mini_NAL}), we first search for the  \civ\  absorption features  at $1300<\lambda_{\rm rest}<1550$ \AA.  
Any troughs with a signal-to-noise ratio (S/N) less than five  in REW are considered unreliable detections.  Four different-type \civ\ troughs, namely the BAL, mini-BAL, NAL, and AAL, are firmly identified via the approach.  
These classifications  are  based  on  visual inspection of  trough width,  using the two thresholds of 2000 \kms\ and 500 \kms\ to distinguish BAL, mini-BAL, and NAL,  while the AAL  displays a triple-peaked trough that is apparently composed of  three \civ\ doublets.   
We model  the four distinct \civ\  troughs  with the  minimum number of Gaussians from the normalized spectrum at MJD=53674, when noticing the model parameter degeneracy  between  spectral resolution and trough saturation (e.g., \citealp{Vietri22,MaoH2025,Perna2025}).  Each of the  \civ\  doublets from the  NAL and AAL  is modeled by two Gaussians,  with a fixed separation of 497 \kms\ and a full width at half maximum (FWHM) tied to each other during the fit.

Noticeably, both the  NAL  and AAL  at MJD=53674 appear to have at least one \civ\ doublet with the blue/red ratio close to 1:1 and FWHM $>300~\kms$, indicative of outflows with a partial coverage and somewhat saturation. Of particular interest is a successive  \civ\  line-locking signature in the  AAL, which reinforces an outflow origin for the AAL,  although the  line-locking physics  could be more complicated  than that  from previous studies based  on a single ion from  low- or intermediate-resolution spectra (e.g.,  \citealp{Simon2010,Lewis2023,ChenC24,ChenC25,LuW25}).  The BAL and mini-BAL, which are not resolved in velocity and likely saturated to some extent,  can be modeled with one broad Gaussian.   
Given the monotonic decrease of trough depth,  we choose to use the \civ\  apparent trough depth (ATD; the deepest portion of each trough; see the blue shadings in Fig.~\ref{J075852_com_BAL_mini_NAL}) at MJD=53674 as a proxy of the effective LOS coverage,   which can be used in assessing the long-term accumulated trends, i.e., chemical dilution and the competition of cloud shredding vs. growing  (\citealp{Fielding2022}),  across free-expansion outflows along our LOS.  Noticeably,  the decreasing ATD trend of \nv\ holds across the mini-BAL, NAL, and AAL from a spectrum normalized to either the continuum+BEL or the pure continuum; furthermore, the ATD ratio of \nv/\civ\ is insensitive to short-term variability in column density and/or covering factor.  Therefore, the trends of ATD across them can be firmly established, regardless of the specific normalization level and short-term variability.  Due to  unidentifiable  \ovi/\hi\ troughs from the BAL, the heavy blending in \nv, and transverse motion (see next Section), the  BAL was not measured in ATD throughout  this work.   
%  .  

\subsubsection{Variability of  the \civ\ BAL    }
\label{BAL_variability} 

BAL variability is widely explained by transverse motion across our LOS, ionization change, or a mixture of both.  
One  can  immediately learn  from Fig.~\ref{J075852_com_BAL_mini_NAL} that the entire \civ\ BAL trough exhibited a monolithic shift in velocity from MJD=53674 to 55568 characteristic of BAL deceleration,  followed by its  disappearance over the next spectroscopic epochs. This monolithic  shift  was quantitatively confirmed by the cross-correlation function (CCF) analysis across the entire BAL trough.  Specifically,  we obtain  a velocity shift of 950$_{-51}^{+126} \kms$  from the CCF analysis via  3000 Monte Carlo simulations, with the error bars depicting the  90\% percentile confidence level  (see \citealt{Yi24} and references therein). 
The fitted \civ-BAL profiles in the first two epochs can be used as a  template to match the corresponding \nv-BAL troughs after scaling in flux (see the dashed  profiles from the \nv-BAL panel of Fig.~\ref{J075852_com_BAL_mini_NAL}),  reinforcing the argument for BAL deceleration.  The physical mechanism in producing  BAL-deceleration signatures could be complicated (\citealp{Yi24}).  Here, we highlight the scenario where an absorber moves along a curved path as invoked by \citet{Gabel2003}.

Alternatively,  the observed velocity shift may be  caused by two BAL absorbers,  given  the obvious asymmetry of the BAL profile   at MJD=55568.   Indeed, we found that the BAL trough at the 1st/2nd epoch can be modeled with one/two broad Gaussians, respectively; in specific,  the one at $v_{\rm LOS}\sim -36000~\kms$  shrinks across the entire trough  over the five  epochs, while the other at $v_{\rm LOS}\sim -33000~\kms$ appears only in the  spectrum at MJD=55568,  supportive of transverse  motion by  two  BAL absorbers.  
Nevertheless, both interpretations are compatible with Keplerain motion, a scenario where the distance can be well constrained with current data (see Section \ref{discuss_density}).

\subsubsection{Variability of the mini-BAL and NAL in \civ\ }
\label{coordinated_var}

Unlike the  BAL exhibiting a large  velocity shift  along with a decrease of ATD  from MJD=53674 to 55568,   the mini-BAL slightly  weakened  while the NAL did not show any  variability in this time interval.   However, the mini-BAL almost vanished  and the NAL weakened  by a factor of $\sim$2  in the second time interval from MJD=53674 to 59517, characteristic of  delayed variability ($\sim$1.2 rest-frame  yr) for the NAL. 
This  variability behavior  offers a unique diagnostic to constrain the recombination timescale (see Section~\ref{discuss_density}).  
Both the  mini-BAL and NAL  weakened monotonically across the entire trough with a commensurate change in ATD over two decades,  favoring  ionization change  (e.g., \citealp{Hamann11,Capellupo2011,Filizak13,Yi19a,HeZ19}),   given that it can act as a common driver of  variability across  different-scale flows. 

This ionization-change scenario is independently supported by an overall increasing diminution of ATD across the \ovi, \nv, and \civ\ troughs over two decades (see  the  mini-BAL or NAL in Fig.~\ref{J075852_com_BAL_mini_NAL}),   perhaps due to nonblack saturation contributing to a lesser extent across these ionic troughs in the same absorber. Furthermore,  ATD can vary due solely to ionization changes without transverse motion in an inhomogeneous environment (e.g., \citealp{Hamann11,Yi2021}),  lending credit to our interpretations of different trough-variability behaviors, i.e.,  transverse motion for the BAL as opposed to ionization change for the mini-BAL and NAL.   
Last but not least, if the (H, He) ionization fronts  exist,  trough variability would be very sensitive to small changes in physical state (\citealp{Gabel2005}), even if the trough is saturated to some extent (see a local analogue with the response of days from \citealp{Kriss19}).

\subsubsection{The lack of variability in the AAL }
The red component of the \civ\ doublet ``a'' embedded in  the AAL trough (see the \civ-AAL panel from Fig.~\ref{J075852_com_BAL_mini_NAL}), at first glance, is dramatically shallower in the first epoch than in the later epochs. However,  this variability  is unlikely to be true,  since the normalized spectrum at MJD=53674 has no sampling data point with the normalized flux below 0.6 at  $400<v_{\rm LOS}<650$ \kms.  In fact,  such apparent variability can be firmly attributed to an  instrumental artifact  and/or  an unsatisfied spectral extraction at this epoch,  especially when noticing strong telluric absorption in this velocity range.  This explanation  is independently  supported by  the lack of significant variability in the corresponding red member of the \siiv\ doublet from the AAL  (see  ``a, b'' from the \civ-BAL panel of  Fig.~\ref{J075852_com_BAL_mini_NAL}),   given the highly correlated variability between  \civ\ and \siiv\ (e.g., \citealp{Filizak13,Yi19a}). Therefore, we  conclude that the \civ\ trough  remains  unchanged in REW over two decades.

Previous studies reported that $\sim$80\% strong \civ\ AALs in quasar spectra likely arise from outflows,  especially those having a broad, complex trough (e.g., \citealp{Misawa07,Simon2010,ChenC18,Rudie2019}),  in the context that  ISM  have been shredded and dispersed by previous blowout events (e.g., \citealp{Hopkins2010,Faucher12}).  
We identified from J0758 a complex feature characterized by  the triple-peaked trough in \civ,  which can be decomposed as (at least) three sets of \civ\ doublets with a fixed velocity separation  of $\approx$430 \kms\ in successive order characteristic of line-locked  outflows (see \citealt{Hamann11} for the case of a ﬂow being not aimed directly at us),  a phenomenon that  largely rules out  chance alignments from (unseen) companion galaxies. On the other hand, the lack of strong Ly$\alpha$ damping wings from the AAL, again, disfavors an origin of neighboring galaxies,  whereas  absorbers outside of the CGM may contribute somewhat to the AAL in \ovi\ and \hi\ (see Section~\ref{multiple_ions}).

\begin{table*}
\centering
 \caption{  Spectral measurements }
 \begin{tabular}{lccccccccc }
  \hline %\noalign{\smallskip}
\civ & REW-1 &  REW-2  &  REW-3  &  $v_{\rm cen}$ & FWHM$_{\rm d}$  & ATD &  $\lambda_{\rm N/C}$  &PCF  &  $\tau_{1551}$     \\
           & (\AA) &  (\AA)  &  (\AA)  &  ($\kms$) &  ($\kms$) &  &  &     \\
  \hline\noalign{\smallskip}
BEL & 12.10$\pm$0.18 & 12.3$\pm$0.16 & 12.4$\pm$0.10 & --2500 & - & - & - & - & -  \\
BAL & 8.37$\pm$0.22 & 6.51$\pm$0.17 & 3.60$\pm$0.10 & --36000 & 5510 & - &  -  & - & -   \\
mini-BAL & 1.89$\pm$0.11 & 1.28$\pm$0.09 & 0.26$\pm$0.07 & --20700 & 1045 & 0.37$\pm$0.05 &  1.70$\pm$0.08 & - & -    \\
NAL & 2.01$\pm$0.09 & 1.92$\pm$0.07 & 1.10$\pm$0.05 &  --11800  & 330 & 0.55$\pm$0.04 &  0.82$\pm$0.06 & 0.56 & 2.8   \\
AAL-a & 2.15$\pm$0.06 & 2.25$\pm$0.05 & 2.20$\pm$0.04 & --70 & 230 & 0.93$\pm$0.03 & $<$0.27 & 1.00 & 1.4   \\
  %\noalign{\smallskip}
  \hline
\end{tabular}
\label{table2}
 \footnotesize{ \\
 Note:  REW-1/2/3 refers to the rest-frame equivalent width (REW) measured from the first/second/third epoch, respectively.  The BEL and AAL REWs are calculated at $1500<\lambda_{\rm rest}<1543$ \AA\ and $1545<\lambda_{\rm rest}<1549$ \AA, respectively.  $v_{\rm cen}$ and ATD refer to the centroid  velocity and apparent trough depth at MJD=53674.  FWHM$_{\rm d}$ is measured from the deepest Gaussian component for each trough.   $\lambda_{\rm N/C}$ is the ATD ratio of \nv/\civ.  PCF is the partial covering factor derived from the velocity revolved \civ\ doublet and $\tau_{1551}$ is the \civ(1551) optical depth derived from the  NAL and  AAL at the deepest point.  
 }
\end{table*}

\subsection{Constraints on density and distance}\label{discuss_density} 

Ideally, the recombination timescale characteristic of a response to changes in the incident ionizing flux can be obtained by high-cadence spectroscopic and photometric observations, if the source is not in the ``holiday'' state  (e.g., \citealt{Kriss19}). 
As pointed out in Section~\ref{coordinated_var}, variability of the mini-BAL and NAL is  likely caused by ionization changes.  Therefore,  the  interval from MJD=53674 to 59517 ($\Delta t=3.63$ rest-frame yr) can be used as an upper limit of recombination timescale and the  interval from MJD=53674 to 55568 ($\Delta t=1.18$ rest-frame yr) as a lower limit of recombination timescale  for the NAL; then, one can further constrain its electron density from the equation under the assumption of the ionization rate lower than the recombination rate 
\begin{equation}
n_{\rm e} \sim (t_{\rm rec}*\alpha_{\rm rec})^{-1}
\end{equation}
where $n_{\rm e}$ is the number density and $\alpha_{\rm rec}$ is the recombination coefficient (we choose $2.45\times 10^{-11}$ cm$^{-3}$~s$^{-1}$ throughout  this work; e.g., \citealp{Hamann11,Rogerson16}).   As a result,  $n_{\rm e}$ is constrained to be $350\lesssim n_e\lesssim1100$   cm$^{-3}$ for the  NAL, consistent with observational studies of kpc-scale, emission-line  outflows (e.g., \citealp{Forster19,XuX2020}).  
Similarly, the first time interval  can be used as an upper limit of  recombination timescale for the mini-BAL, so its density is measured to be $n_{\rm e}>1100$  cm$^{-3}$, consistent with \citet{Misawa14} where they inferred an upper limit on distance of $\sim$kpc.

When it comes to the  AAL that remains unchanged in REW,  one has to distinguish the saturation effect, i.e., column density would vary in response to changes in ionizing flux but without exhibiting significant variability in REW (see Fig.22 in \citealt{Yi19a} for demonstration),  from the low-density effect, i.e., a slow response to changes in ionizing flux. The former scenario is unlikely given (1) the lack of variability in the unsaturated \siiv\ from the AAL-a,  and (2) the \civ(1551) optical depth of 1.4 at the deepest point of the AAL-a suggestive of a largely unsaturated trough.  
In the latter scenario, the time interval  from MJD=53674 to 60786 sets a  lower limit on recombination timescale for the AAL,  then its  density is   estimated to be $n_{\rm e}< 290$  cm$^{-3}$. 
On the other hand, the highest-S/N spectrum at MJD=59529 reveals weak absorption in \sii\ 1260\AA\ and \cii\ 1335\AA\  (see the AAL-Ly$\alpha$ panel of  Fig.~\ref{J075852_com_BAL_mini_NAL}).  Given the spectral quality, it is reasonable to set the \sii\ ground/excited ratio $\gtrsim$5,  then electron density is constrained to be $n_{\rm e}\lesssim$175 according to the Eq.5 from \citet{Hamann01},  consistent with that from our variability analysis. 
Therefore, both results  strongly favor a small electron density and hence a large distance.  
On the other hand, if the AAL is close to the nucleus, one would expect to detect variability due to transverse motion like the BAL, regardless of nonblack saturation. 
Indeed, the AAL  density  falls into the ``responds too slowly'' region at a minimum distance of $R>$1.4 kpc from a lower-luminosity quasar (see Fig.14 from \citealt{Rogerson16}), which naturally explains the stable AAL over two decades.  With $L$(1500\AA)=8.3$\times10^{46}$ erg s$^{-1}$, $n_{\rm e}\lesssim175$ and $U=0.012$ (see Section \ref{PI_analyses}), its distance can be constrained at $R\gtrsim$25 kpc, according to the Eq.10 from \citet{ChenC18};  furthermore, if the \cii\ ground/excited ratio  $\gtrsim$5, the AAL would have $n_{\rm e}\lesssim$49 cm$^{-3}$ and hence  $R\gtrsim$48 kpc from the quasar, shedding light on the surprisingly high fraction of unbound multiphase absorbers at $z\sim$2 (\citealp{Rudie2019}). 
 As a comparison, \citet{Perrotta2018} predicted from a more sophisticated modeling  the opposite trends of column density  between \nv\ and \civ, as well as the presence of  low-ion absorption features like \sii\ on a  scale of $100\lesssim R\lesssim$1000 kpc (see Fig.8 in their work),  consistent with what we observed in J0758.    
Similarly, the NAL ($U$=0.35; if $n_{\rm e}$=500) and mini-BAL ($U$=0.55; if $n_{\rm e}$=10000) distances are estimated to be $R\sim$2.8 and $R\sim$0.5 kpc, respectively.

Conversely, the BAL is the only one among the four absorption outflows exhibiting a monolithic velocity shift.  In the context of Keplerian motion, the BAL distance is estimated to be $R\lesssim$16 pc from the quasar, under the assumption of an BAL absorber with the LOS coverage of 0.25 crossing a homogeneous background light source with a size of 0.01 pc (e.g., \citealp{Rogerson16,Yi2022}),  indicative of the innermost one among the four type outflows.  This distance is  in line with that from another luminous quasar at $z\sim3$, which exhibits a wider and stronger EHVO that may be crucial for quasar feedback (\citealt{Paola25}).

To summary,  the distance order across these outflows can be firmly established,   paving the way for our analysis of radial gradients.

\subsection{ A  joint analysis with other ionic troughs   }
\label{multiple_ions}

To gain a better  understanding of the absorption,  we  display  the wavelength regions where other prominent ions would be present at the same velocities for the BAL, mini-BAL,  NAL, and AAL.  The  \civ\ trough of interest  was used as a benchmark to search for other ionic troughs in the same velocity range.  
We  chose to display in Fig.~\ref{J075852_com_BAL_mini_NAL} for the corresponding \nv,  \ovi, and \hi\ Ly$\alpha$  troughs that aligned  with the four distinct \civ\ troughs in velocity, although the identification of  Ly$\alpha$  is more difficult than the other two (singlet vs. doublet).  
Indeed, the apparently high \civ/Ly$\alpha$ ratio favors an outflow origin for the AAL; furthermore, a good match of the ``a,b,c'' in velocity from the AAL between Ly$\alpha$ and \civ,  adds independent support to the scenario where three outflowing absorbers are line locked and hence make an almost full LOS coverage. 
In this context, the triple-peaked AAL trough can be naturally explained by the swept-up ISM/CGM via three outflowing absorbers, which becomes line locked  by the overwhelming quasar radiation in a later time (e.g, \citealp{WangT2024,ChenC24,ChenC25}). 
Due to unidentifiable BAL troughs in \ovi\ and \hi, however, we do not perform such analyses for the innermost BAL undergoing transverse motion (but see \citealt{Paola25} for a detailed analysis of an EHVO like this BAL).  

\begin{figure*}
    \centering
    \includegraphics[width=0.33\linewidth]{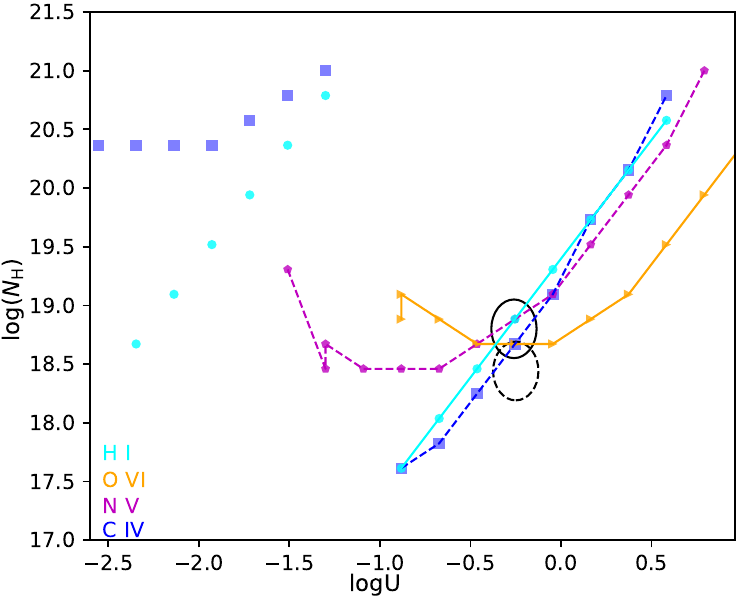}
    \includegraphics[width=0.33\linewidth]{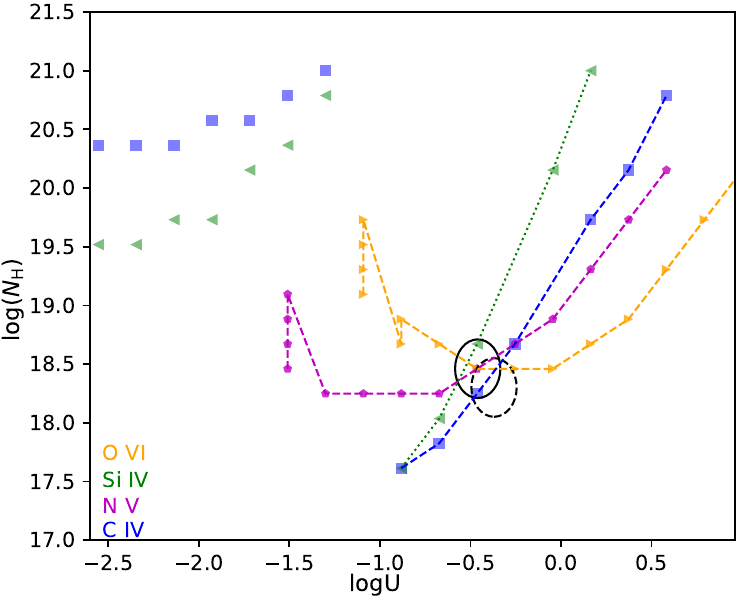}
    \includegraphics[width=0.33\linewidth]{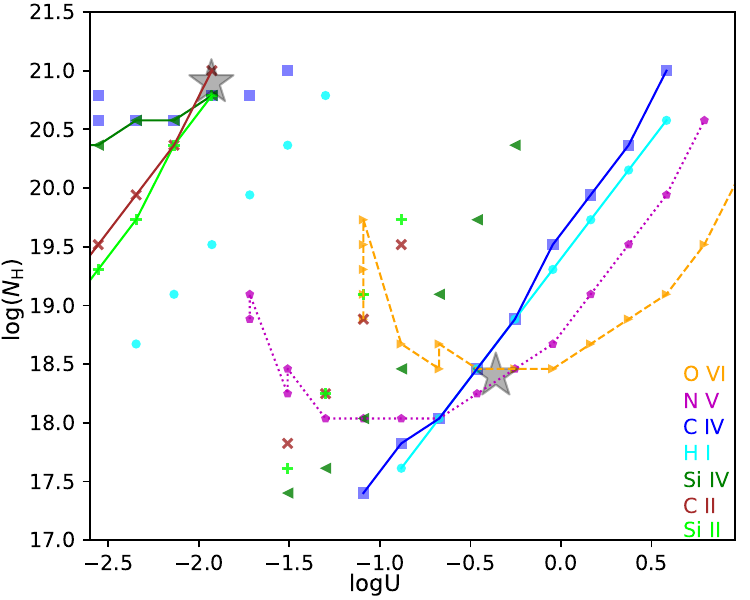}
    \caption{
      Left: The CLOUDY solutions for the variable mini-BAL from the 1st/3rd  (solid/dashed ellipses) epoch. Middle: The CLOUDY solutions for the variable NAL from the 1st/3rd epoch. Right: The two-zone solutions (pentagons) estimated from photoionization modeling via CLOUDY for the multiphase AAL that remains unchanged in REW over two decades. The dotted/dashed curves in each panel refer to the upper/lower limit on the ionic column density, respectively, while  solid curves depict the relatively well constrained ions. 
    }
    \label{J075852cloudy_for_AAL_NAL}
\end{figure*}

Our observations reveal  a remarkable  gradient of the \nv/\civ\  ratio ($\lambda_{\rm N/C}$) derived from the opposite ATD trends between \nv\ and \civ\ across the mini-BAL, NAL, and AAL,  which hold at each epoch regardless of the normalization level and short-term  variability, making the gradient of $\lambda_{\rm N/C}$ tenable.  Furthermore,   the  turning point of a shallower trough in \nv\ than in \civ\ ($\lambda_{\rm N/C}<1$)  appears to be first present in the NAL,  consistent with a free expansion transitioning to energy-driven outflows on kpc scales (\citealt{Costa20}). The outermost AAL exhibits  a much  shallower  trough in \nv\ than in \civ,  a phenomenon that  could be related to the wind-CGM coupling  (see Section \ref{schematic_model}); furthermore,  the AAL ionization state may be no longer dominated by the quasar (e.g., \citealp{Tumlinson2017,Chisholm2018,LiuW22}), a phenomenon reminiscent of the positive correlation between ionization parameter and projected distance from CGM to intergalactic scales (\citealt{Lau2016}). This scenario is supported by (1) the absence/presence of a monotonic ATD trend across the mini-BAL, NAL, and AAL in Ly$\alpha$/\civ, respectively,  (2) the stable  AAL as opposed to dramatic variability in the other troughs, and (3) an upper limit on $\lambda_{\rm N/C}$ from the AAL and an increasing \siiv\ ATD from the NAL to AAL characteristic of a huge dilution via swept-up CGM (\citealt{Simcoe2006}).   The multiphase nature evidenced by the coexistence of \sii, \cii, \siiv, \civ, \nv, and \ovi\ in the AAL,  which is commonly seen from galactic winds or CGM absorption, may trace a large-scale turbulent cascade from outflows coupling to the CGM (e.g., \citealp{Fielding2022,Sameer2024,QuZ2025}).

\subsubsection{ Effective covering factors and column densities }
\label{CF_ionic_column}
 
Intrinsic absorbers often partially cover the background light source, as indicated by the relative strength of the blue/red members in a doublet. Thus, one may need to distinguish the covering factor from optical depth effects before the measurement of column density, although covering factor can vary due to ionization change without transverse motion (see Section \ref{coordinated_var}). Following a similar prescription from the literature (e.g., \citealp{Gabel2005, Kara2021}) and ignoring the narrow-emission contribution (given the weak \oiii\  emission; see Section \ref{BH_Edd}),  our  analyses reveal that all but the AAL cover  only a part of the continuum source (see Fig.~\ref{J075852_3spec_com}),   indicative of nonblack saturation.

We adopt the apparent optical depth (AOD; \citealp{Tripp1996}) method to place the lower limit on column density, while the upper limit is constrained from partial covering for the  \civ\ troughs, with an additional assumption of free of bending and contamination for the \ovi\ and \nv\ troughs.  All the measured column densities are listed in Table \ref{table3}, which in turn provide us a benchmark to search for photoionization solutions.

\subsubsection{ Solutions from photoionization modeling  }
\label{PI_analyses}

To further explore the mini-BAL, NAL and AAL,  we performed  photoionization modeling via CLOUDY (e.g., \citealp{Chatzikos2023,ChenC25}),  assuming a solar relative abundance pattern, a photoionization equilibrium, and a typical AGN ionizing SED (with temperature of 20000 $K$, $\alpha$(ox)=--1.6, $\alpha$(uv)=--0.5, $\alpha$(x)=--0.9), across the grid range from --3.6 to 1.0 with 0.2 dex steps in ionization parameter ($U$) and from 17.4 to 21.0 with 0.2 dex steps in hydrogen column density ($N_{\rm H}$). As shown in Fig.~\ref{J075852cloudy_for_AAL_NAL}, the solutions across the mini-BAL, NAL and AAL in the 1st epoch (with log$U$=[--0.26, --0.46, --0.36\&--1.93] and log$N_{\rm H}$=[18.8, 18.46, 18.4\&20.9]) exists in the case with log$N$(\civ) = [14.67, 14.69, 14.9] and log$N$(\nv) = [14.88, 14.67, 14.5], which also agree with the observed trends of ATD across the three outflows, despite challenges from nonblack saturation. 

\begin{table}
\centering
 \caption{  Ionic column densities measured from the 1st/3rd epoch }
 \begin{tabular}{lccr}
  \hline\noalign{\smallskip}
Ion    & mini-BAL  & NAL  & AAL-a \\
     & (log)  & (log)  & (log) \\
\hline\noalign{\smallskip}
\ovi\  & $>$15.02  & [15.04, 15.62] &  $>$15.0 \\
         & $>$14.49   & [14.72, 15.30] &  $>$15.0 \\
\nv\   & $>$14.88  &  [14.67, 15.47] &  14.2$\pm$0.7 \\
         & $>$14.34  &  [14.50, 15.08] &  14.2$\pm$0.7 \\
\civ\  &  $>$14.67 & [14.69, 15.25] &  15.0$\pm$0.7 \\
         &  $>$14.15  &  [14.43, 15.25] &   15.0$\pm$0.7 \\
\siiv\ &  -           &   [13.28, 13.86] & 14.3$\pm$0.7 \\
         &  -           &   $<$13.0 &14.3$\pm$0.7 \\
  \hline
\hi(1216)  &  $<$15.1  &    -      &  14.5$\pm$0.7 \\
\cii(1335)   &  -           &    -        & 13.5$\pm$0.7 \\
\sii(1260)  &  -  &    -        & 12.5$\pm$0.7 \\
 \noalign{\smallskip}\hline
\end{tabular}
\label{table3}
\end{table}

However, some caveats must be kept in mind during the analyses. 
The AAL characterized by strong \ovi\ and Ly$\alpha$ troughs may suffer from blending due to the nearby intergalactic medium (IGM) and/or a mixture of cooling/heating,  given the weakest absorption in \nv\ (with an intermediate ionization potential) relative to \ovi\ and \civ\ (e.g., \citealp{Simcoe2006,Chisholm2018}). 
Since the spectral quality becomes increasingly noisy toward higher-order \hi\ Lyman series at the spectral blue end,  the strong  \ovi\ and  \hi\ troughs cannot be accurately decomposed with current data.  Importantly, our assumptions of the same metallicity ($Z$) and ionizing SED for all the three outflows are overly simplified and may not hold in reality.   The differences can reach 0.7 dex and 0.9 dex when using different metallicity and ionizing SED, respectively; furthermore, the derived $U$ may vary by $\sim$1.5 dex when a trough is more sophisticatedly decomposed into individual components from high-resolution spectroscopy (e.g., \citealp{Rudie2019,MaoH2025,Perna2025}). 
On the other hand, simultaneously accounting for the high-ionization (\ovi\ and \nv) and low-ionization (\sii\ and \cii) species in a single cloud is notoriously difficult, as hinted by the large difference in trough width between \sii(1260) and \ovi,  although this issue may be alleviated by a coherent physical picture (\citealp{QuZ2025}). In fact,  at least  two phases are needed to match the observed measurements for the AAL (see Fig~\ref{J075852cloudy_for_AAL_NAL}).  
Therefore, the photoionization solutions here should be viewed as exploratory rather than conclusive. Nevertheless, the distances can be further constrained from photoionization modeling compared to that from variability analyses (see Section \ref{discuss_density}), which in turn shed light on the physical  implications (see Section \ref{schematic_model}).

\section{The origin of weak \oiii\ emission and the central engine properties} \label{BH_Edd}

 Apparently,   the near-IR spectrum shows weak or negligible \oiii\ emission,   consistent with the presence of a strong outflow  and/or metal cooling in the narrow emission-line region (NLR;  e.g., \citealp{Matsuoka2017,Yi20}), especially when noticing the high-velocity NAL located at a scale ($R\sim$1 kpc) overlapping with the quasar NLR. Interestingly, the increasing \siiv\ or Ly$\alpha$ ATD from the NAL to the AAL, to some extend, implies that the \siiv\ and \hi\ gas may cluster at $R>1$ kpc, where outflows start to cool and mix with ISM.  From a technical view, however, the weak \oiii\ emission in J0758 may be caused by the long-slit (with a slit width of 1\arcsec ) spectroscopic effect,  when noticing the spatially extended \oiii\ emission from a mini-BAL quasar at $z=3.5$ (\citealp{Perna2025}).  The co-existence of the absorption outflows, which acts as a natural coronagraph, offers a unique opportunity to probe the spatial extension using an integrated field spectrograph in the future.

\begin{figure}
    \centering
    \includegraphics[width=0.95\linewidth]{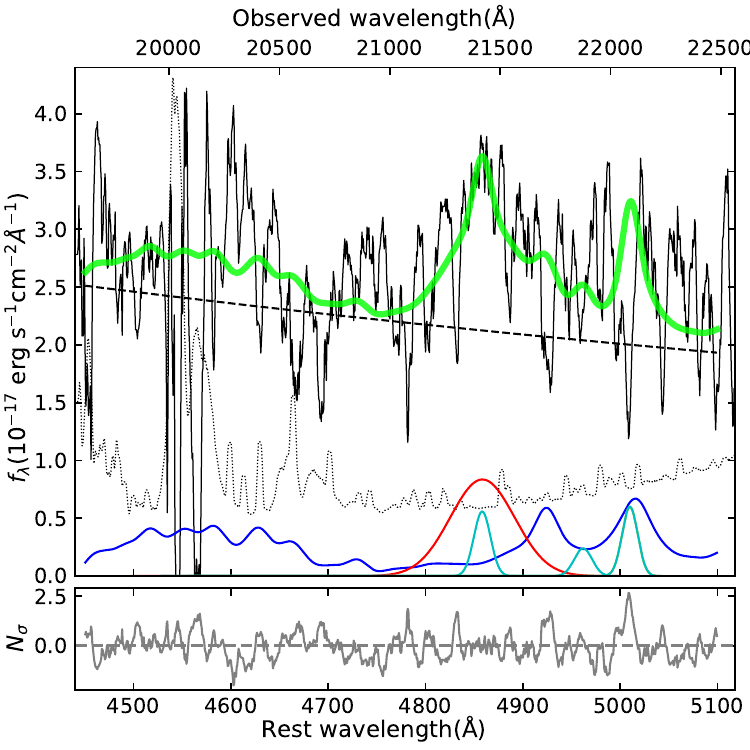}
    \caption{
      Spectral decomposition for the complex emission around the H$\beta$ emission.   The  broad H$\beta$ (FWHM = 4780 $\kms$) and  narrow components (FWHM = 1240 $\kms$)   are displayed in red and cyan lines, respectively; the dashed line indicates the continuum  and the blue line refers to the \feii\ component;  the total fit is shown in green. $N_\sigma$ in the bottom  is the residual. 
    }
    \label{J0758HbFit}
\end{figure}

 Since the broad  H$\beta$ emission line is thought to be the most reliable tracer of supermassive black hole (SMBH) mass in quasars,  we attempt to decompose the complex emission line around  H$\beta$ following the same prescription of \cite{Yi20},  despite the noisy near-IR spectrum.  
 As shown in Fig.~\ref{J0758HbFit},  the broad  H$\beta$ component is measured to be $\sim$4780 $\kms$.  Then, we estimate the SMBH mass of $\sim$4$\times 10^9 M_\odot$ and Eddington ratio of $\sim$0.83,  with the uncertainty dominated by the systematic error (0.3 dex) from the single-epoch scaling relation.  Our estimates  support  that the presence of such a luminous quasar with multiscale winds is driven by an SMBH accreting at the Eddington limit, which in turn lends support to  the  weak \oiii\ emission,  although it may be  affected by the low-S/N, long-slit spectrum.  

\begin{figure*}
        \centering % \flushleft
        \includegraphics[width=.67\textwidth]{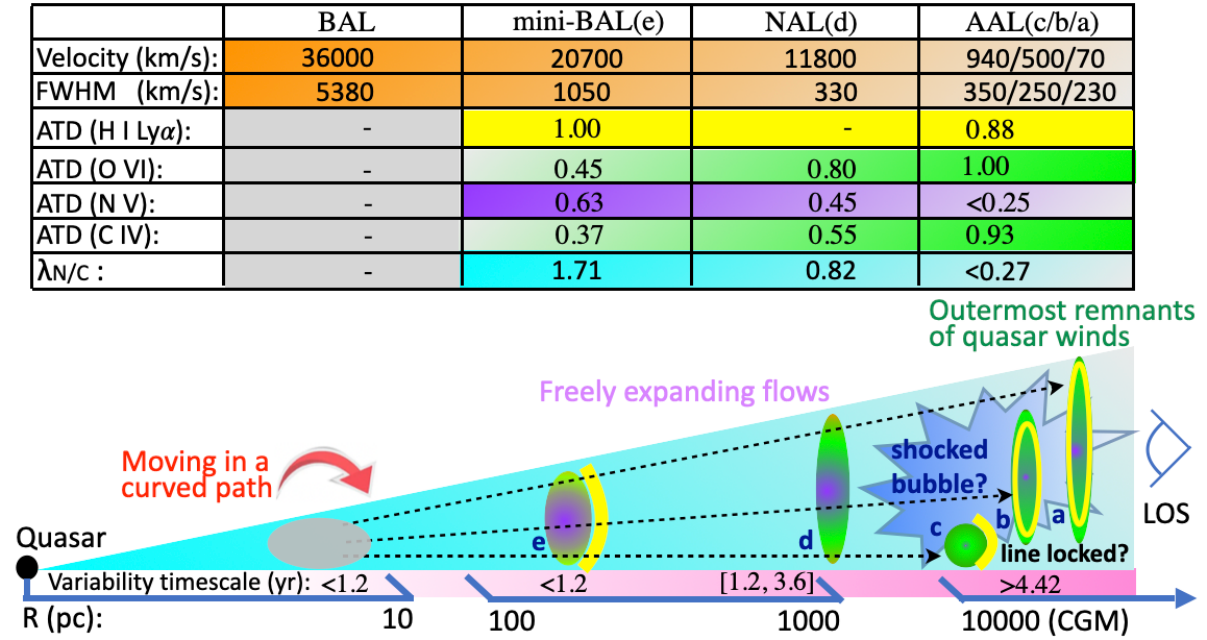}
    \caption{   A   cartoon illustration of the expansion for J0758  from a side view,  in which the characteristic trough depth/width  in \civ\ monotonically increases/decreases with  decreasing  velocity across these outflows.  Variability timescale increases with distance,  characteristic of an increasingly  diluted gas density  (see text).  The innermost  BAL   moved  out of our LOS  along  a curved  path,   causing a large velocity shift,  followed by  the  disappearing BAL and mini-BAL,  as well as the weakening NAL with a time delay of $\sim$5 (1.2 rest-frame) yr;   in contrast,   the AAL is the outermost remnant of quasar winds and remains unchanged in REW over two decades.  The presence of  monotonic trends  across these  outflows favors a coherent picture:  the mini-BAL and NAL are freely expanding outflows, perhaps due to the outermost AAL clearing away substantial ISM along our LOS;  the  NAL is the first one  showing  $\lambda_{\rm N/C}<1$ among them,  signaling a transition from high-N/C, ejective feedback on small scales to low-N/C,  regulative feedback on large scales, potentially along with shocked bubbles responsible for  strong \hi\ Ly$\alpha$ absorption detected in the mini-BAL and AAL (e.g., \citealp{Adelberger2003,Simcoe2006}).  }
    \label{snow_plowing_model}
\end{figure*}

\section{Discussion}\label{schematic_model}
 
Based on these distances constrained above,  the BAL, the mini-BAL,  the NAL, and the AAL in J0758 are most likely located  in ascending order  from nuclear to CGM scales.  
Here,  we propose  a snowplowing picture for the coexistence of these  outflows,  i.e., they  are multiscale and perhaps episodic quasar winds,  whose  velocity, electron density, and trough width overall decrease  during the expansion (see Fig.~\ref{snow_plowing_model}).   
Therefore,  we are likely witnessing the long-sought transition from a  high-velocity  nuclear wind  to a low-velocity  outflow  on  kpc scales  (e.g., \citealp{King11,Faucher12,Costa20,Veilleux22,Hall2024}).  This scenario  naturally explains the stable AAL  as opposed to dramatic  variability of the BAL and mini-BAL  over two decades,   in the context that the outermost AAL already travelled to a large-scale ($R>$10 kpc) region and coupled to the CGM, leading to a diluted outflow density and a  mixture of  ionizing/cooling processes (e.g., \citealp{Simcoe2006,Fielding2022,LiJ2026}).

Strong \nv, especially compared to \civ, has been rarely detected on a scale larger than that of the host galaxy (e.g., \citealp{Simon2010,WuJ2010,Lau2016,Perrotta2016}).  
Noticeably,  $\lambda_{\rm N/C}$ was first detected  below unity in the NAL and dropped to less than 0.27 in the AAL (see Fig.~\ref{J075852_com_BAL_mini_NAL} and \ref{snow_plowing_model}),  indicative of an expansion from the quasar host to its CGM.     
Together with other characteristic gradients (e.g., a similar trend of trough width in \citealp{Simon2010}), we interpret them  as the signpost of being caught on a critical  transition from ejective  feedback on circumnuclear scales to regulative  feedback on circumgalactic scales, along with a mixture of ionizing/cooling processes and putative shocks (e.g., \citealp{Jannuzi96,Adelberger2003,Forster19,Nelson2019,Costa20,Veilleux20,WangT2024}),  in the context that \nv\ arises from a compact region and has been transported to $\sim$kpc scales via outflows before encountering the ISM, while \civ\ and \ovi\ mostly trace the homogeneous halo gas (e.g., \citealp{Hamann1999,Prochaska2014}). Intriguingly, such a transition explains the large difference in the \nv\ detection rate along and across the LOS (down-the-barrel vs. transverse directions; e.g., \citealp{Perrotta2016,Lau2016}),  although  in-depth analyses are needed to explore the wind-ISM/CGM coupling physics.

Both  the  \civ\ BAL and mini-BAL disappeared over two decades;  in contrast,  the NAL and  AAL  are still persistent  after the BAL disappearance, providing a clue of the formation of complex  AALs among the non-BAL quasars. Perhaps more  importantly,  the aforementioned transition from free-expansion to energy-driven outflows can exert  a far-reaching impact to its host-galaxy evolution,  since energy-driven outflows accompanying shocked bubbles would persist a long time after the demise of quasar ($\gtrsim$10 times quasar lifetime; e.g., \citealp{King11,Lochhaas2018,Nelson2019,Costa20}), and retain most of nitrogen within the host galaxy due to the (inverse) shocks,  making a long duration of (relatively) high-velocity, multiphase galactic outflows possible, i.e., the AAL  in J0758,  other AAL outflows from the literature (e.g., \citealp{Adelberger2003,ChenC18,LiuW22,Yi24}),  the large-scale, emission-line outflows (e.g., \citealp{Forster19,XuX2020}) etc.  This scenario can  also explain  the prevalence  of  large-scale   absorption/emission  outflows characteristic of relic AGN-driven winds observed in massive galaxies without apparent AGNs (e.g., \citealp{Simcoe2006,Tremonti2007,Forster19}).   
There are  two  competing mechanisms  (quasar-driven vs.  starburst-driven),  by which  high-velocity galactic outflows  can be  launched  from a compact region (e.g., \citealp{Nelson2019}).   J0758 provides a bona fide example of quasar-driven outflows spanning from nuclear to CGM scales,  complementing the long-sought blowout picture (see an early-stage case from \citealt{Yi2022} and a late-stage case from \citealt{Hamann01} at cosmic noon). This scenario is also in line with the multi-stage or episodic feedback models (e.g., \citealp{Hopkins2010,LiJ2026}),  given an increase of ATD in \civ\ (or \ovi) as opposed to a decrease of electron density from the inner to outer winds characteristic of a snowplowing expansion.  The origin of \hi\ gas and its relation to these outflows, however, remain puzzling given the absence of any trends in this ion.

\section{Summary and Future Outlook }
\label{discussion_sec}

It has long been speculated that there is an underlying link between small-scale winds and large-scale outflows, i.e.,  high-velocity NALs are relic  quasar winds in massive galaxies  with or without ongoing quasar activities (e.g., \citealp{Hamann01,Kuraszkiewicz2002,Adelberger2003,Simcoe2006,Tremonti2007,Misawa07,Veilleux2017,ChenC20,FuX2023}), despite the lack of solid observational evidence.  The coexistence of multiscale outflows in J0758, however,  offers a unique opportunity to test the underlying link with  low cost.  Specifically, the gas kinematics, the short-term variability behaviors,  and the long-term trends  observed from down-the-barrel spectroscopy of J0758,  for the first time to our knowledge,  signify a spectacular expansion from nuclear to CGM scales  (see Fig.~\ref{snow_plowing_model}).  
$\lambda_{\rm N/C}$, which is insensitive to short-term variability, can serve as a unique probe to investigate the long-term, accumulated effects during the expansion driven by multiscale winds. The stable and perhaps line-locked AAL likely traces the outermost remnant of quasar winds coupled to CGM,  resulting in the multiphase structure that is commonly seen from galactic winds (e.g., \citealp{Tumlinson2017,Rudie2019,Veilleux20,QuZ2025}).

A more comprehensive picture of the wind-ISM/CGM coupling in J0758 can be substantially achieved by multi-wavelength, high-spatial/-spectral resolution  observations and/or sophisticated simulations, i.e.,  spatially resolved spectroscopy would be critical for investigating the relation between emission/absorption outflows, and high-spectral resolution spectroscopy with detailed ionization modeling may offer additional tools to probe the inner physics. Future observations and simulations may be also helpful to further assess the long-term effects and provide valuable diagnostics to explore the intricately intertwined ``ecosystem'' that has  plagued the broad AGN-galaxy community for many decades.

\section*{Acknowledgements}

C.J.W. and W.Y. acknowledge the support from National Key R\&D Program of China No.2022YFF0711500, 2023YFA1608300, the support from ``Yunnan Revitalization Talent Support Program'' - Young Talent Project: Research on the Key Technologies of the Observation Control System Based on Data Driven, and the National Natural Science Foundation of China  (NSFC-11703076). 
P.R.H and E.R.P acknowledge support from the National Science Foundation AAG Award AST-2107960.  
P.R.H and E.R.P acknowledge support from the National Science Foundation AAG Award AST-2107960.   P.B.H. is supported by NSERC grant 2023-05068.  C. C. is supported by NSFC-12103097. Z. C. He is supported by the NSFC (Grant Nos. 12222304, 12192220, and 12192221). X.-B. Wu acknowledges the financial support from the NSFC (Grant No.12133001).  K.X.L. acknowledges financial support from the NSFC-12573020, the Young Talent Project of Yunnan Province,  and the Youth Innovation Promotion Association of Chinese Academy of Sciences (2022058).

Many thanks to the referee for valuable comments and suggestions. 
We thank Lei Hao and Hongyan Zhou for stimulating discussions.   We are grateful to the help from Weijian Guo for obtaining the DESI data and Huarui Bai for the data reduction of the LJT/YFOSC data.  
This research  uses data obtained through the Telescope Access Program (TAP), which has been funded by the National Astronomical Observatories of China, the Chinese Academy of Sciences (the Strategic Priority Research Program ''The Emergence of Cosmological Structures'' grant No. XDB09000000), and the Special Fund for Astronomy from the Ministry of Finance. 
CC thank the support of  the National Natural Science Foundation of China (12073097). 
Observations obtained with the Hale Telescope at Palomar Observatory were obtained as part of an agreement between the National Astronomical Observatories, the Chinese Academy of Sciences, and the California Institute of Technology. 
The Hobby-Eberly Telescope (HET) is a joint project of the University of Texas at Austin, the Pennsylvania State University, Ludwig-Maximillians-Universit\"{a}t M\"{u}nchen, and Georg-August-Universit\"{a}t G\"{o}ttingen. The Hobby-Eberly Telescope is named in honour of its principal benefactors, William P. Hobby and Robert E. Eberly. 
The Low-Resolution Spectrograph 2 (LRS2) was developed and funded by the
University of Texas at Austin McDonald Observatory and Department of
Astronomy, and by the Pennsylvania State University. We thank the
Leibniz-Institut f\"ur Astrophysik Potsdam and the Institut f\"ur
Astrophysik G\"ottingen for their contributions to the construction
of the integral field units. 
Funding for SDSS-III has been provided by the Alfred P. Sloan Foundation, the Participating Institutions, the National Science Foundation, and the U.S. Department of Energy Office of Science. 
The DESI Legacy Imaging Surveys consist of three individual and complementary projects: the Dark Energy Camera Legacy Survey (DECaLS), the Beijing–Arizona Sky Survey (BASS), and the Mayall z-band Legacy Survey (MzLS). 
We acknowledge the support of the staff of the Lijiang 2.4 m telescope (LJT). Funding for the telescope has been provided by CAS and the People’s Government of Yunnan Province.



%\bibliographystyle{mnras}
%\bibliography{oao1657}

\end{document}